\documentclass{iau} 
\usepackage{graphicx}
\usepackage{cite}

\title[Feedback in Galaxy Outflows] 
{AGN and Star Formation Feedback in Galaxy Outflows}

\author[E. de Gouveia Dal Pino, W. Clavijo-Boh\'{o}rquez  \& C. Melioli]   
{Elisabete de Gouveia Dal Pino$^1$, William Clavijo-Boh\'{o}rquez$^1$ and Claudio Melioli$^2$
}

\affiliation{$^1$Instituto de Astronomia, Geof\'isica e Ci\^encias Atmosf\'ericas  (IAG), Universidade de  S\~{a}o Paulo, \\ 
CEP 05508-090, S\~{a}o Paulo, Brazil \\ email: {\tt dalpino@iag.usp.br, weclavijob@usp.br} \\[\affilskip]
$^2$University of Modena, Italy 
}

\pubyear{2018}
\volume{342}  
\jname{Perseus in Sicily: from black hole to cluster outskirts}
\editors{Keiichi Asada, Elisabete de Gouveia Dal Pino, \\Hiroshi Nagai, Rodrigo Nemmen, Marcello Giroletti, eds.}
\begin{document}

\maketitle

\begin{abstract}
Large-scale, broad outflows are common in active galaxies. In systems where star formation coexists with an AGN, it is unclear
yet the role that both play on driving the outflows.
In this work we present three-dimensional radiative-cooling MHD simulations of the formation of these outflows, considering  the 
feedback from both the AGN and  supernovae-driven winds.
We find that a large-opening-angle AGN wind develops  fountain structures that make the expanding  gas  to fallback. Furthermore, 
it exhausts the gas near the nuclear region, extinguishing star formation and accretion within a few 100.000 yr, which establishes
the duty cycle of these outflows. 
The AGN wind accounts for the highest speed features in the outflow with velocities around 10.000 km s$^{-1}$ (as observed in UFOs),
but these are not as cold and dense as required by observations of molecular outflows. The SNe-driven wind is the main responsible
for the observed mass-loading of the outflows. 

\end{abstract}

\firstsection 
\section{Introduction}
Fast massive, poorly collimated  outflows of HI and molecular gas, with velocities $\sim$  600-1500 km s$^{-1}$, have been recently observed  in the
central regions of AGN sources like  Seyfert galaxies and ULIRGs (e.g, Morganti et al 2005, 2013; see also R. Morganti in these procs.). Also, much hotter 
ultra-fast outflows (UFOs) with velocities $\sim$ 0.03-0.3$c$  have been detected in X-rays (e.g. Tombesi et al. 2013). 
An interesting example of a molecular outflow is the
 ULIRG AGN 4C12.50. It exhibits HI and  CO gas outflow with velocity $\sim$ 1000 km s$^{-1}$ which is co-spatial  with the radio jet hot-spot, at 
 100 pc scales (Morganti et al. 2013, 2014, Dasyra and Combes 2012). This is probably produced by radiative cooling of shocked gas due to 
 jet interaction  with the ISM. Typically, gas column densities $\sim 10^{21}-10^{22}$ cm$^{-2}$ are inferred for these outflows,  mass 
 loss rates up to 100 M$_{\odot}$ yr$^{-1}$, and HI clumps with mass larger than 600 M$_{\odot}$.
 A spectacular example of UFO is the one detected (in X-rays) in the  Seyfert NGC 4151 central region, denominated Eye of Sauron, which is 
 surrounded by ionized hydrogen (HII) associated to star formation regions observed in optical and is also connected with infalling HI gas 
 observed in radio with VLA (Wang et al. 2010). UFOs have even larger column densities $\sim 10^{23} - 10^{24}$ cm$^{-2}$ 
 (Kraemer et al. 2018). 
 
The investigation of these outflows  is important as they transport energy and gas outwards from the central regions and may affect the
evolution of the host galaxy and its surrounds as well. 
Their origin is not yet fully understood, but the accreting supermassive black hole  (SMBH) in the center of the AGN is believed to be 
their main driving source. Injection mechanisms such as  radiation pressure, or the kinetic interaction of the radio plasma with the ISM, 
or even  magneto-centrifugal processes are often invoked (e.g. Kremer et al 2018 and references therein). A key point when investigating  these outflows is the fact that they 
are generally associated with Seyferts, for which SF regions and  starbursts (SB)  coexist with the AGN, and the energy power emitted by both 
the nuclear AGN and the host galaxy is comparable, so that it is unclear the relative role of the SF and the AGN on the driving  of the 
outflows (Melioli and de Gouveia Dal Pino 2015). 

In this work, we explore the formation and evolution of these outflows in a Seyfert-like galaxy  (within 1 kpc scale) by means of 
3D MHD simulations at high resolution (3.9 pc) including gas radiative cooling,  magnetic fields and the mechanical feedback from 
both SF (through the production of type Ia and II SNe) and AGN outflows.

\section{Results of the MHD Simulations}

For the numerical setup of our simulations, we have considered a total gravitational potential provided by dark matter, the  galactic bulge
and the disk. We adopted a black hole to bulge mass ratio $10^{-3}$ and considered a multi-phase, initially stratified  gas disk supported 
both  by rotation, and  thermal and magnetic pressures. We have considered a SF rate of 1 M$_{\odot}$ yr$^{-1}$,  type Ia SNe rate
$=0.01$ yr$^{-1}$ 
and  type II SNe rate $0.1$ yr$^{-1}$. We have considered equilibrium radiative cooling for a gas able to achieve temperatures  between  $T=100$ K to $10^8$  K. 
We have taken initial thermal to magnetic pressures $\beta \equiv \infty$, $=$300, 30, 3,  column densities $10^{22}$  to $10^{23}$  cm$^{-2}$,   
AGN wind initial aperture angles $= 0^{\circ}$, $10^{\circ}$, or spherical,  with injection power $10^{43}$ erg s$^{-1}$,  velocity $v = 0.06c$ and 
mass rate $5 \times 10^{-4}$ M$_{\odot}$ yr$^{-1}$. (For more details see Melioli and de Gouveia Dal Pino 2015; Clavijo-Boh\'{o}rquez, de Gouveia Dal Pino and Melioli, in prep.).

As an example, Figure  compares four evolved models at $t= 5$ Myr. These diagrams clearly indicate that a larger AGN outflow opening angle has 
a major influence on the nuclear region evolution, sweeping the ISM matter, dragging the SN-driven winds, quenching accretion and star formation, 
and accelerating much more gas to velocities above 10.000 km s$^{-1}$. The comparison of the results for the two different initial magnetic field 
geometries do not reveal substantial differences, this in part because of the large value adopted for $\beta$ here (which is compatible to the 
values expected from observations at these scales). Nevertheless, the comparison of the velocity-density diagrams with pure HD simulations 
indicate that the presence of $B$ helps to preserve and increase the number of denser structures at higher (positive and negative) velocities.


Our  simulations also show that the winds driven by the products of SF alone (i.e., by explosions of supernovae, SNe) can drive outflows 
with velocities  $\sim 100-1000$ km s$^{-1}$ and  mass outflow rates of the order of 50 M$_{\odot}$ yr$^{-1}$ (Figure  , 
which resemble the properties of warm absorbers (WAs) and molecular outflows. However, the resulting densities ($\sim 1-10$ cm$^{-3}$) and 
temperatures (between $10^4$ and $10^ 5$ K) are too low and too high, respectively, compared to the expected values of these outflows.


\begin{figure}
\begin{center}
 \includegraphics[width=5.2in]{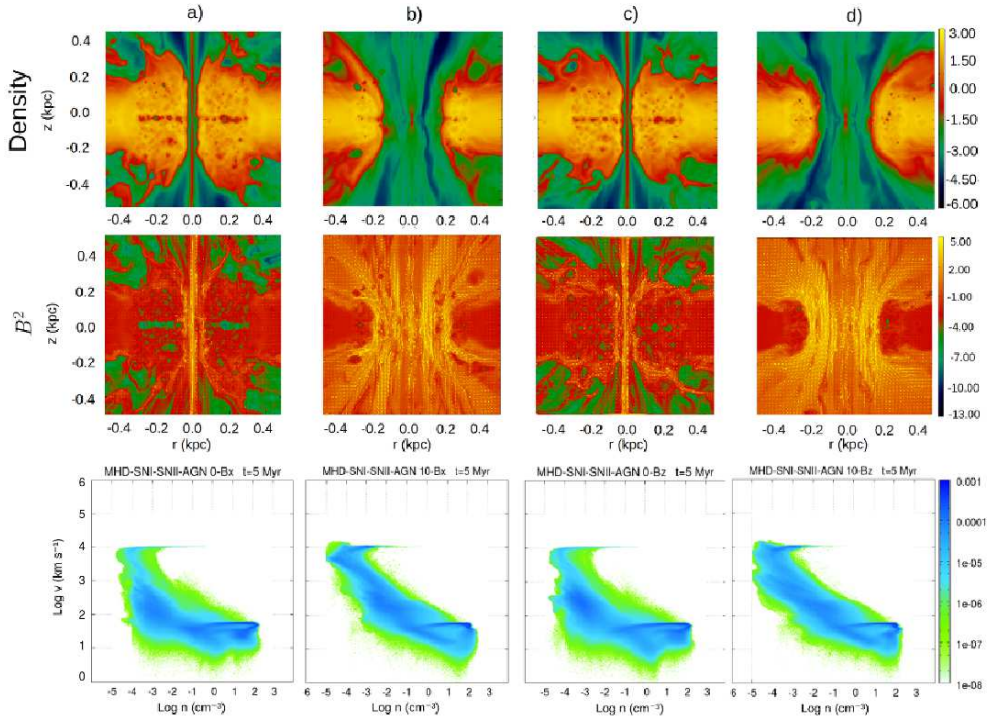} 
 \caption{Top: Two-dimensional (2D) central cuts of the density distribution within one-kpc region of the evolved galaxy (at 5 Myr), for four 
different initial conditions. From left to right: (a) a system with initial homogeneous horizontal magnetic field ($\beta =$ 300, and 
$B_x=0.76 \mu$G) and an AGN outflow with $0^{\circ}$ opening angle (collimated); (b) the same as in the left panel, but with the AGN outflow 
with $10^{\circ}$ opening angle; (c) the same as in (a), but with initial homogeneous vertical magnetic field ($B_z$); and (d) the same as in (c),
but with the AGN outflow with $10^{\circ}$ opening angle. Middle: 2D central cuts of the magnetic field strength ($B^2$), with the magnetic field 
vectors superposed to it. Bottom: 2D histograms of the vertical velocity versus density distribution, calculated considering all the cells of
the systems. Velocity is in km s$^{-1}$ and density in cm$^{-3}$, both in logarithmic scale. The color bar indicates the cell numbers normalized to 
their total number (see more details in Clavijo-Boh\'{o}rquez, de Gouveia Dal Pino, Melioli in prep.). }
\label{density}
\end{center}
\end{figure}


\begin{figure}
\begin{center}
 \includegraphics[width=5.3in]{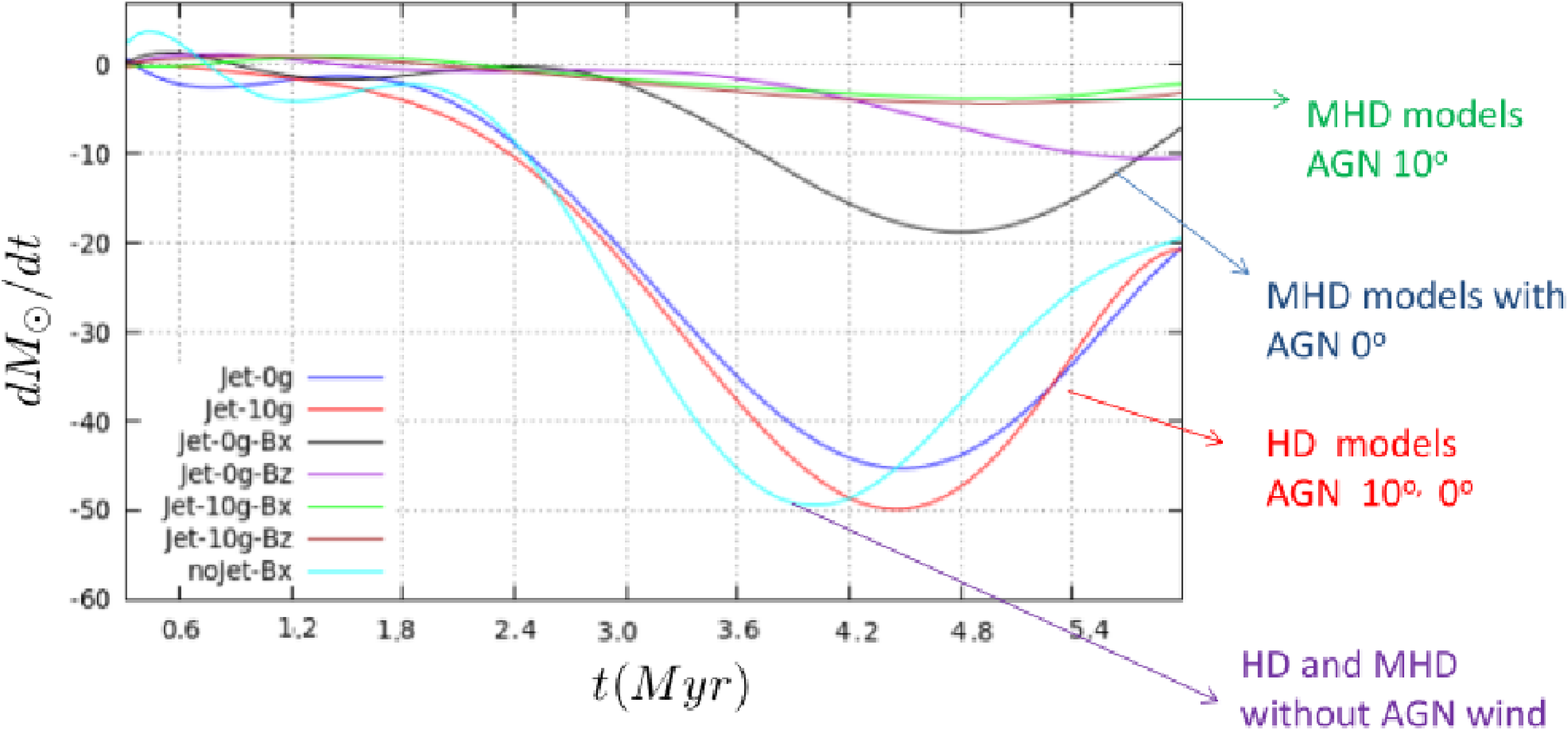} 
 \caption{Time evolution of total  gas  mass loss rate from the  thick	disk of the galaxy ($|z| \leq 200$ pc) for different models as indicated by the colors (see more details in Clavijo-Boh\'{o}rquez, de Gouveia Dal Pino, Melioli in prep.).}
   \label{mass-core-dmdt}
\end{center}
\end{figure}

\begin{figure}
\begin{center}
 \includegraphics[width=5.4in]{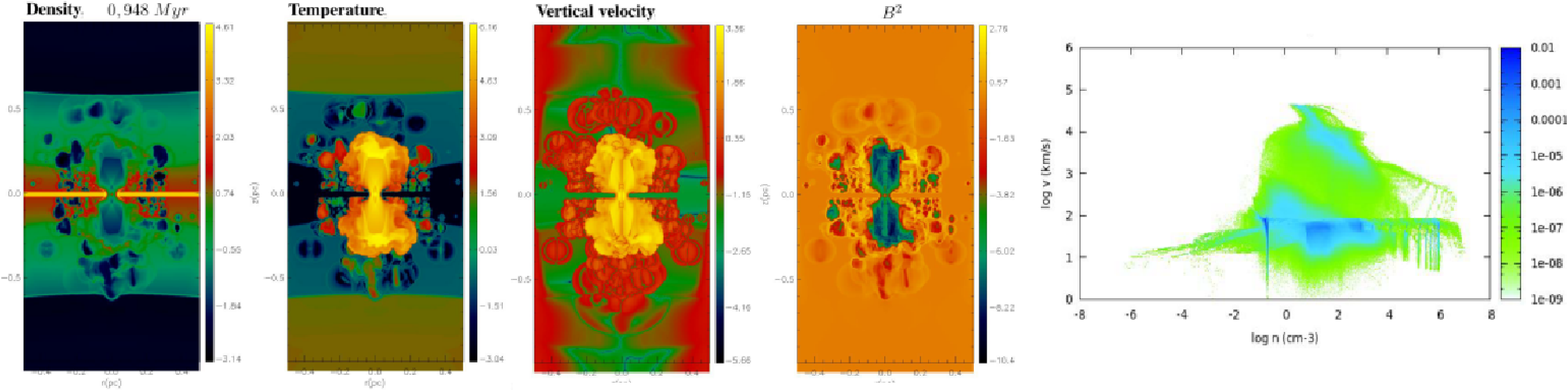} 
 \caption{Early evolution for a  model with an AGN wind injected spherically in the nuclear region with initial power  $2 \times 10^{45}$  erg s$^{-1}$, mass rate $1.5$ M$_{\odot}$ yr$^{-1}$, and velocity   
$0.21$ c. From left to right: 2D cuts of density, temperature, velocity and magnetic field strength distributions; and velocity-density histogram as in Figure \ref{density} (see more details in Clavijo-Boh\'{o}rquez, de Gouveia Dal Pino, Melioli in prep.).}
   \label{early-jet}
\end{center}
\end{figure}


The presence of an AGN-wind, particularly  with  large opening angle, causes the formation of fountain-like structures (Fig. \ref{density}) 
that make part of the expanding gas (pushed also by the SF-wind) to fallback, producing a "positive" feedback on the host galaxy evolution. This 
effect is more pronounced in presence of magnetic fields, due to the action of extra magnetic forces  by the AGN. This reduces the mass loss rate in the outflows by factors 
up to 10 when compared with HD models (see Figure \ref{mass-core-dmdt}).

Figure 3 depicts the early evolution ($t \sim$ 950.000 yr) for  a model with an  AGN-wind injected  spherically in the nucleus,  after the SNe-driven wind is entirely developed. 
We note that the wind is quickly collimated by the surrounding gas forming dipolar outflows that resemble  UFOs.     
The diagrams and the density-velocity histogram show that high speed, dense structures with velocities larger than  10.000 km s$^{-1}$, temperatures $>$ 10$^7$ K, and  densities $>$ 100
cm$^{-3}$ do develop. Besides, intermediate  speed structures are also detected with velocities 
$\sim$ 1000 km s$^{-1}$, temperatures $>$ 1000 K, and  densities $<$ 10$^6$ cm$^{-3}$.


\section{Discussion and Conclusions}

MHD radiative cooling simulations of the nuclear region (1 kpc$^3$ scale) of Seyfert-like  galaxies,  taking into account  both the SF and  AGN outflow feedback, result  the following:

The SNe can drive outflows with velocities of a few 1000 km/s, but the characteristic densities are too low and the temperatures too large in comparison to the observations of WA and molecular outflows.  

A collimated AGN jet  alone (without SNe-driven wind) is  unable to drive massive outflows, but can  accelerate structures to very high speeds (as observed in UFOs).

An AGN wind with large opening angle sweeps more ISM matter, accelerates more  gas to $v >$ 10.000 km s$^{-1}$, but it exhausts gas fuel and quenches accretion onto the SMBH and star formation 
in the nuclear regions within a few  $100.000$ yr. This indicates that  the duty cycle of these outflows is around a few 100.000 yr, which is compatible with the timescales  inferred for the observed UFOs and molecular outflows.

Mass loss rates up to 50 M$_{\odot}$ yr$^{-1}$ (which are compatible with the observations of  molecular outflows) are driven mainly by the SNe-driven-wind component.

Large opening angle AGN-winds, specially in MHD flows, favor the fallback of gas to the galaxy (fountain) which decreases the  mass loss rate.

An AGN spherical wind injected when the SNe wind is fully developed  improves the results, producing  UFO characteristics   ($v>$ 10.000 km s$^{-1}$, $T>$ 10$^7$ K, and $n >$ 100 cm$^{-3}$), 
but  the slower outflow  component with $v \sim$ 1000 km s$^{-1}$ is too hot to explain the molecular outflows.

Missing ingredient in our models, like the presence of non-equilibrium ionization radiative cooling
to account for more efficient coolers (molecules and dust) that could survive to evaporation and destruction in interactions  with high speed hot gas, will help to improve our results and  
 ensure the survival of the cold, dense clumpy structures that are swept to very high speeds as observed in  molecular outflows. 

\section{Acknowledgments}
We acknowledge support from the Brazilian agencies FAPESP (2013/10559-5 grant) and CNPq (306598/2009-4 grant). The simulations presented in this lecture have made use of the computing facilities of the GAPAE group (IAG-USP) and the Laboratory of Astroinformatics IAG/USP, NAT/Unicsul (FAPESP grant 2009/54006-4).


\begin{thebibliography}{}

\providecommand{\natexlab}[1]{#1}
\providecommand{\url}[1]{\texttt{#1}}
\expandafter\ifx\csname urlstyle\endcsname\relax
  \providecommand{\doi}[1]{doi: #1}\else
  \providecommand{\doi}{doi: \begingroup \urlstyle{rm}\Url}\fi
  
    \bibitem[{Dasyra} \& {Combes}(2012)]{2012A&A...541L...7D}
{Dasyra, K. M \&  Combes, F.} 2012, \textit{A$\&$A}, 541, L7.

\bibitem[Amari \etal\ (1995)]{Amari_etal95}
{Amari, S., Hoppe, P., Zinner, E., \& Lewis R.S.} 1995, \textit{ApJ}, 30, 490


\bibitem[Kraemer \etal\ (2018)]{2018ApJ...852...35K}
{Kraemer, S. B.,  Tombesi, F., \& Bottorff M. C.} 2018, \textit{ApJ}, 852, 35

  
\bibitem[Melioli \& de Gouveia Dal Pino (2015)]{2015ApJ...812...90M}
{Melioli, C., \&  de Gouveia Dal Pino, E. M.} 2015, \textit{ApJ}, 812, 90


\bibitem[Morganti \etal\ (2005)]{2005A&A...439..521M}
{Morganti, R., Oosterloo, T. A., Tadhunter, C. N., van Moorsel, G., \& Emonts, B.}, 2005 
\textit{A$\&$A}, 439, 521


\bibitem[Morganti \etal\ (2013)]{2013A&A...552L...4M}
{Morganti, R., Frieswijk, W., Oonk, R. J. B., Oosterloo, T., \& Tadhunter, C.} 2013,
\textit{A$\&$A}, 552, L4

\bibitem[Morganti \etal\ (2015)]{2015ASPC..499..125M}
{Morganti, R., Oosterloo, T. A., Oonk, J. B. R., Frieswijk, W., \& Tadhunter, C. N.} 2015, \textit{ASP-CS}, 499, 125
  
\bibitem[Oosterloo \etal\ (2000)]{2000AJ....119.2085O}
{Oosterloo, T. A., Morganti, R., Tzioumis, A., Reynolds, J., King, E., McCulloch, P., \& Tsvetanov, Z.} 2000, \textit{AJ}, 119, 2085
  

\bibitem[Tadhunter \etal\ (2014)]{2014Natur.511..440T}
{Tadhunter, C., Morganti, R., Rose, M., Oonk, J. B. R., \& Oosterloo, T.} 2014, \textit{Nature}, 511, 440

\bibitem[Tombesi \etal\ (2011)]{2011ApJ...742...44T}
{Tombesi, F., Cappi, M., Reeves, J. N., Palumbo, G. G. C., Braito, V., \& Dadina, M.} 2011, \textit{ApJ}, 742, 44

\bibitem[Tombesi \etal\ (2013)]{2013MNRAS.430.1102T}
{Tombesi, F., Cappi, M., Reeves, J. N., Nemmen, R. S., Braito, V., Gaspari, M., \& Reynolds, S.} 2013,
\textit{MNRAS}, 430, 1102


\bibitem[Tombesi \etal\ (2015)]{2015Natur.519..436T}
{Tombesi, F., Mel{\'e}ndez, M., Veilleux, S., Reeves, J.M., Gonz{\'a}lez-Alfonso, C., \& Reynolds, C. S.} 2015, \textit{Nature}, 519, 436


\bibitem[Wagner \etal\ (2013)]{2013ApJ...763L..18W}
{Wagner, A. Y., Umemura, M., \& Bicknell, G. V.} 2013, \textit{ApJ}, 763, L18

\bibitem[Wang \etal\ (2010)]{2041-8205-719-2-L208}
{Wang, J., Fabbiano, G., Risaliti, G., Elvis, M., Mundell, C. G., Dumas, G., Schinnerer, E., \& Zezas, A.} 2010, \textit{ApJ}, 719, L208


\end{thebibliography}
\end{document}